# Terahertz Emission from Spintronic Stack Nanodecorated with Drop-Cast Core–Shell Plasmonic Nanoparticles

Vittorio Cecconi[1,2], Akash Dominic Thomas[1], Ji Tong Wang[1], Cheng-Han Lin[1], Anoop Dhoot[1], Antonio Cutrona[1], Abhishek Paul[1], Luke Peters[1], Luana Olivieri[1], Elchin Isgandarov[1], Juan Sebastian Totero Gongora[1], Alessia Pasquazi[1] and Marco Peccianti[1]

[1] Emergent Photonics Research Centre, Department of Physics, School of Science, Loughborough University, LE11 3TU, UK
[2] DIET, Sapienza University of Rome, Via Eudossiana 18, 00184 Rome, Italy

Spintronic emitters promise to revolutionise terahertz (THz) sources by converting ultrafast optical pulses into broadband THz radiation without phase-matching constraints. Because the conversion relies on spin-current injection across a nanometre-thin magnetic layer, its efficiency is ordinarily limited by weak optical coupling. Here, we present a demonstration of a drop-casting based approach to introduce ultrafast plasmonic-mediated coupling: a sparse-layer of silica–gold core–shell nanoparticles is deposited directly onto a W/Fe/Pt spintronic trilayer. This sparse (≈ 6 %) decoration increases the wafer-averaged THz pulse energy, pointing to a very high local conversion enhancement for this low-coverage spintronic emitter compared with the bare stack. This demonstration points to a viable pathway toward highly efficient spintronic terahertz emitters with potential applications in spectroscopy, imaging, and ultrafast technologies.

## Introduction

Terahertz (THz) radiation (spanning approximately 0.1 THz and 10 THz[1]) has become increasingly pervasive across scientific and industrial domains, with applications in communications[2,3], sensing and non-destructive testing across several sectors.[4]

Progress in these domains is ultimately tied to the development of efficient and scalable THz sources. State-of-the-art approaches, based on various optical-to-THz conversion mechanisms, generally present distinct trade-offs in efficiency, bandwidth, and platform compatibility[1,4].
Photoconductive emitters, characterised by high brightness[5] and modest optical excitation requirements, currently dominate THz spectroscopy applications but do not offer a route for surface scalability or broadband emission. Benchmark nonlinear emitters are based on transparent optical crystals such as ZnTe, $LiNbO_3$, and DAST and generate THz pulses primarily via optical rectification[6]. Owing to their high saturation threshold, they can achieve very high pulse energies and represent a standard technological deployment. However, these remain fundamentally bulk systems, with limits in integration, scalability and bandwidth, as spectral performance is constrained by phase-matching requirements and by phonon-driven absorption (e.g., Reststrahlen band). As an alternative option, semiconductor surface emitters, e.g. low-bandgap media like InAs and InSb, rely on carrier-driven emission mechanisms, enabling high conversion efficiency per unit thickness[7]. Compared to transparent emitters, they benefit from established fabrication processes in the electronics domain, allowing low-cost area scalability. However, they offer low pump saturation[8] and low THz transparency with absorption dominated by Drude free-carrier loss.

Spintronic terahertz emitters (STEs) represent a recent pivotal evolution in overcoming the scalability limits of bulk THz generation technologies. Based on simple nanoscale multilayer structures, they offer phase-matching-free broadband spectral emission with essentially no gaps[9-12]. Over the past decade, their adoption has grown rapidly, accompanied by intense efforts to understand and control the underlying spintronic physics[13-16]. THz radiation from spintronic emitters originates from the excitation of spin-polarised hot electrons in a ferromagnetic (FM) layer by an ultrafast laser pulse. In the most common operating mechanism, the optical excitation promotes majority-spin carriers into higher-energy, more delocalised orbitals, generating a diffusive spin current[17-19]. This current is then injected into a non-magnetic (NM) metallic layer, where it is converted into a transverse charge current via a spin–charge conversion mechanism[20,21], such as the inverse spin Hall effect (ISHE)[22].

In a typical high-efficiency embodiment, the emitter consists of a sandwich structure where theFM layer is embedded between two NM layers with large (and opposite) spin Hall angles (e.g., tungsten and platinum). The structure operates under an external static magnetic field that orients and saturates the FM magnetisation[9,22]. A significant research effort has focused on the dependence of THz emission on the FM and NM layer thicknesses and interfacial properties as optimisation parameters[20] and has shown that STEs can reach emission efficiencies comparable to benchmark nonlinear crystals (e.g., ZnTe)[9]. Despite these advances, optical–FM coupling remains fundamentally limited by the trade-off imposed by the limited diffusion lengths of spin currents, with optimisation often requiring demanding sub-nm control of layer growth[23,24]. Such ultrathin active regions inherently limit the energy transfer from the excitation[25].

To address the limited optical-to-THz conversion efficiency, plasmonic-mediated light coupling has been proposed as a route to enhance the chain of physical mechanisms involved in THz generation. Explored solutions include the use of Au nanoparticles fabricated via high-temperature annealing of an ultrathin Au layer. These particles are subsequently embedded within the ferromagnetic (FM) layer[26], enabling the surface plasmons to act as spin-pump[27]. On the other hand, the idea of introducing field-localisation via engineered optical resonators by surface sparse deposition has been seminally explored by Liu and co-workers[28], who demonstrated THz emission enhancement through the deposition of a bulk layer of gold anisotropic nanorods atop a spintronic trilayer. These approaches tend to be scalable and focus on engineering plasmonic nanoresonators to maximise the local field and its coupling from the surface of the spintronic stack.

More broadly, a large variety of surface nanostructures have been explored for nonlinear responses. These include absorptive metasurfaces[29-31], as well as metallic and plasmonic resonators[32-37], and surface-deposited distribution of resonant nanoparticles, which allows effective tailoring of optical resonances and enhanced local field intensities[38]. Within this context, sparse or self-assembled nanostructures offer an appealing route that combines field enhancement with scalable and straightforward fabrication strategies[32,38-41]. For example, silicon surfaces textured into random "black silicon" needle arrays have shown significantly stronger THz emission under ultrafast optical excitation compared to unstructured substrates[42].

In this work (Fig. 1a), we introduce an ultrafast plasmonic-heating strategy to push spintronic terahertz emitters decisively beyond the established optical–FM coupling limit.

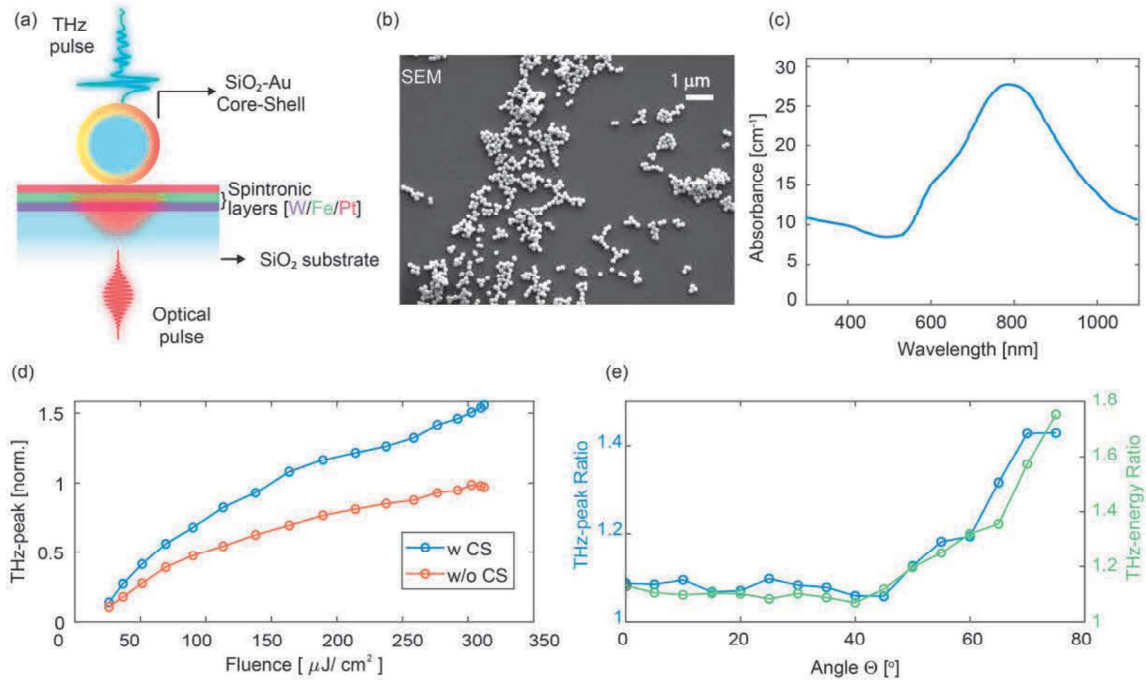

Figure 1. Illustration of the structure and operating principle of the plasmonically-enhanced spintronic assembly. (a) Schematic of the plasmonically-enhanced THz spintronic emitter (STE), consisting of a core–shell nanoparticle positioned on a spintronic trilayer structure composed of W/Fe/Pt, each layer 2 nm thick. (b) SEM image of the silica-gold core-shell nanoparticle deposited on the spintronic. (c) Optical properties of the solution of the core-shell nanoparticles with a peak light absorption around 800 nm (data from nanoComposix Inc. for particles delivered in a water solution); the gold-shell silica-core structure has a diameter of 154±7 nm. (d) Experimental THz emission enhancement as a function of pump power with (blue) and without (red) CS nanoparticles sample at an incidence angle of Θ=75° - data are normalised relative to the reference THz peak-value at the pump peak fluence ~316 µJ/cm$^2$ (390 mW). (e) THz enhancement (peak fields and total pulse energy, with and without CS) vs angle of incidence at fixed power of 390 mW.

By drop-casting a sparse monolayer of Au–SiO$_2$ core–shell nanoparticles (CS) directly onto a nanometre-thick W/Fe/Pt trilayer deposited on a silica substrate, we exploit localised surface-plasmon resonances[43] to funnel pump energy into nanometric hot spots projected onto a spin-polarised magnetic core.

Fig. 1b shows a typical SEM (a full SEM diagnostics is proposed in the supplementary material) image we used to estimate the area coverage of the particles exhibiting monolayer clusterisation and very modest percolation. The average coverage is fully determined by the density of the initial solution.

In our experiment, an average field enhancement from 1.1x to 1.6× as the impinging angle increases towards Θ=75° on the glass substrate (Fig. 1d), with an overall area coverage of 6% is observed, for a fixed pump fluence. While clusterisation makes the local enhancement not homogeneous, the locality of the excitation suggests a strong overall local field-emission enhancement (from about 2.5x to 11× from simple geometrical scaling at the surfaces). Interestingly, this is done without exerting any complex control of the deposition, beyond the solution density.

We highlight that the comparison is proposed at constant fluence and not at constant power, as this reduces the change of temperature (which notoriously affects the spintronic emission efficiency) as we change the angle, while the power at larger angles is consequently higher.

This result also offers context to previous explorations[28] where similar enhancement is observed with anisotropic plasmonic nanorods with 100% surface coverage. In our case, we deploy rotationally-invariant plasmonic resonators, i.e., the local alignment of the particle is not required for the overall effect, and that for this geometry, the enhancement remains robust even with significant clusterisation (see Figure 2).

## Results

The CS nanoparticles act as nanoscale resonators[44], enhancing the local optical intensity, projecting a mode-hot spots that overlap with the W/Fe/Pt spintronic trilayer (Fig. 1a,b). Figure 1b presents a scanning electron microscope (SEM) image of the spintronic surface after deposition, clearly revealing a sparse distribution. The optical response of the nanostructure is resonant at our Ti:Sa pump central wavelength (800 nm). Each core–shell particle has a total diameter of 150 nm, consisting of a silica core surrounded by a 20 nm-thick gold shell. A typical advantage of the CS geometry is the ability to exhibit peaked resonant enhancement in the near-IR, compared to solid nanoparticles[45]. The larger size is beneficial for direct deposition, as it reduces percolation (being also safer to handle and process). The resonance bandwidth accessible with this design is typically compatible with 100 fs-scale ultrafast laser excitations.

In the exploited structure, the core mechanism of spintronic THz emission involves the inverse spin Hall effect (ISHE)[46], which converts optically-injected spin current into a charge current, with a local current density $J_c$ generally governed by

$$J_c = \theta_{SH}(J_s \times \sigma), \quad (1)$$

where $\theta_{SH}$ is the spin Hall angle (a measure of spin-orbit coupling strength), $J_s$ is the spin current, and $\sigma$ is the spin polarisation direction. The charge current is orthogonal to both the spin current and the spin polarisation vector, determined by the magnetic polarisation of the FM medium. The generated charge current acts as a time-varying dipole, which radiates electromagnetic waves in the THz frequency range. Model-wise, the emitted THz electric field $E_{THz}$ is proportional to the time derivative of the charge current

$$E_{THz}(t) \propto \frac{\partial J_c(t)}{\partial t}. \tag{2}$$

The optical coupling of the CS-layer with the spintronic trilayer is strongly influenced by the spatial configuration of core–shell nanoparticles. We numerically simulated how particle clustering induces a spectral shift and broadening in the scattering cross-section for a sample illuminated through the glass substrate. For the core-nanoshell, our numerical estimation shows enhancement that remains effective across various configurations (Fig. 2). The corresponding electric field distributions at 800 nm for different morphology (Fig. 2a-c) reveal that that in general the local field profile is shaped by both the interaction with the impinging light (normal in Fig. 2a-c) and near-field coupling between adjacent nanoparticles, although localisation always occur. Considering moderate clustering (e.g., seven particles, consistent with the general observation in SEM images across our samples), the scattering cross-section around the pumping wavelength is effective in several configurations (Fig. 2d).

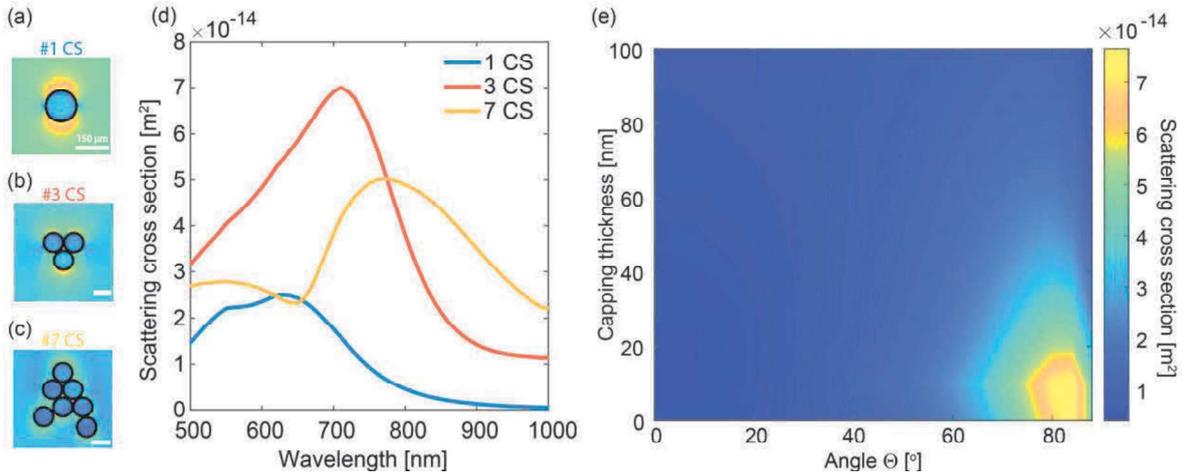

Figure 2. Simulated scattering cross section results and capping thickness study, for a sample illuminated through the glass substrate (a-c). The field value at the middle plane of the core-shell. The black lines represent the boundaries of nanoparticles to show their location. (d) Nanoparticle-induced scattering cross-section of a nanometric platinum layer for three configurations: a single particle, a three-particle system, and a cluster of seven nanoparticles. The results show that for the seven-particle case, the spectrum of the trilayer spintronic broadens, indicating an overall enhancement in tunability. (e) Scattering cross-section efficiency as a function of silica capping thickness and angle of incidence for a p-polarised wave.

We also assessed the role of the distance between the core–shell nanoparticles and the spintronic trilayer in achieving optimal coupling, specifically by introducing a dielectric capping layer as a spacer. Figure 2e shows how varying the thickness of a silica capping layer affects the scattering cross section of the CS nanoparticles. The analysis reveals that increasing the capping thickness leads to a pronounced reduction in the scattering cross section, particularly for p-polarised waves at high incidence angles (Fig. 2e), supporting our design choice to omit any capping to maximise coupling. Notably, the application of nanometric capping layers is widespread in ultrathin device engineering, typically for environmental protection[47,48]. However, the topmost layer in our spintronic stack is platinum, which is resistant to environmental degradation.

In Figure 3a, our numerical calculation predicts a pronounced angular dependence in the absorption of the optical pump within the spintronic trilayer when placed in direct contact with the CS layer. For p-polarised light, absorption is markedly enhanced at large incidence angles, which in turn leads to increased THz emission. Simulated electric field distributions at the spintronic interface further confirm the superior field confinement achieved for p-polarised light at high incidence angles, in contrast to the s-polarised case (Fig. 3b).

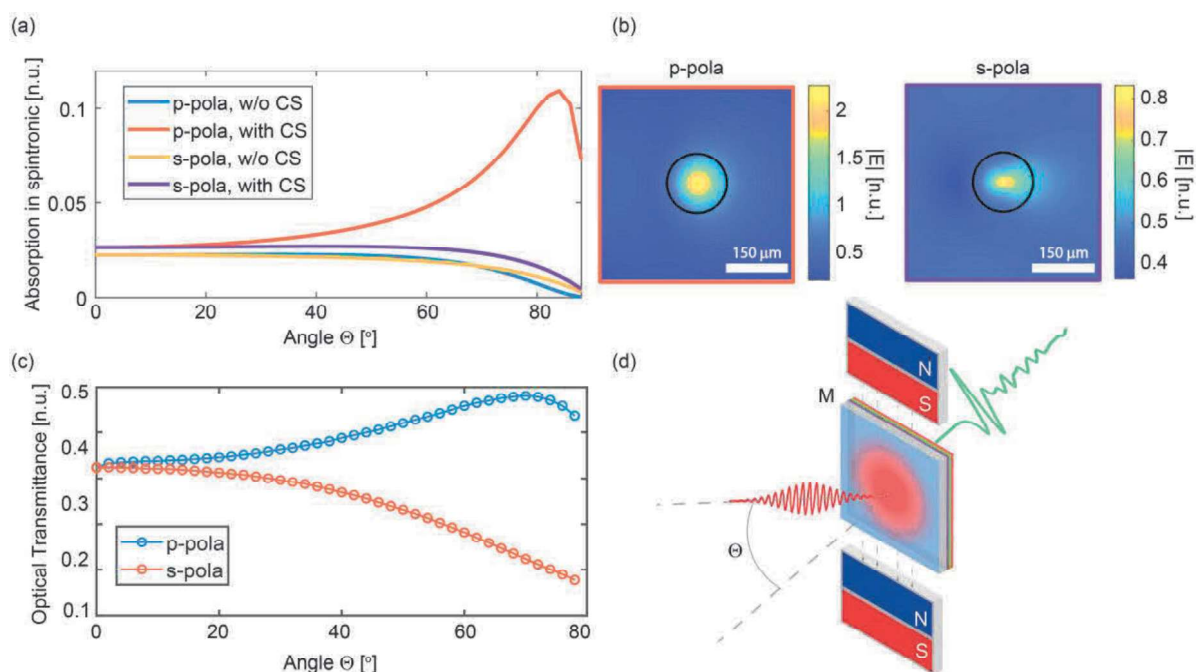

Figure 3. Numerical and experimental optical response of CS nanoparticles (a) Simulated absorption of the optical pump in the trilayer spintronic structure as a function of incidence angle. (b) Simulated electric field distributions at the spintronic plane for p- and s-polarised light at an incidence angle of 84°. The black circle denotes the area where the absorption is calculated in (a). (c) Experimental optical transmission spectra of the spintronic sample at various angles of incidence on the glass substrate, comparing s- and p-polarisations. (d) Schematic of the experimental setup. The sample is immersed in a magnetic field M and rotated in the orthogonal plane, with an angle Θ between the beam and the sample surface.

Notably, Fig. 3c shows the experimental measurements of optical transmission of the spintronic stack across both polarisations. In particular, it shows that the overall transmission of the p-polarisation increases at high angles, both due to the overall reduction of the absorption and impedance matching and corresponds general reduction of the local conversion efficiency.

The measurement setup (Fig. 3d) enables a systematic investigation of the angular dependence of THz emission by rotating the sample within a controlled static magnetic field (~100 mT), with the field direction orthogonal to the rotation axis (see Supplementary Information for schematic details). In this configuration, the sample is illuminated by a Ti:Sa regenerative amplifier (800 nm, 1 kHz, 76 fs pulses, 9 mm $1/e^2$ beam diameter), while the incidence angle Θ between the laser and the sample surface is varied over a wide range (see Supplementary Information for further details). The THz field amplitude is then recorded as a function of Θ (Fig. 1e).

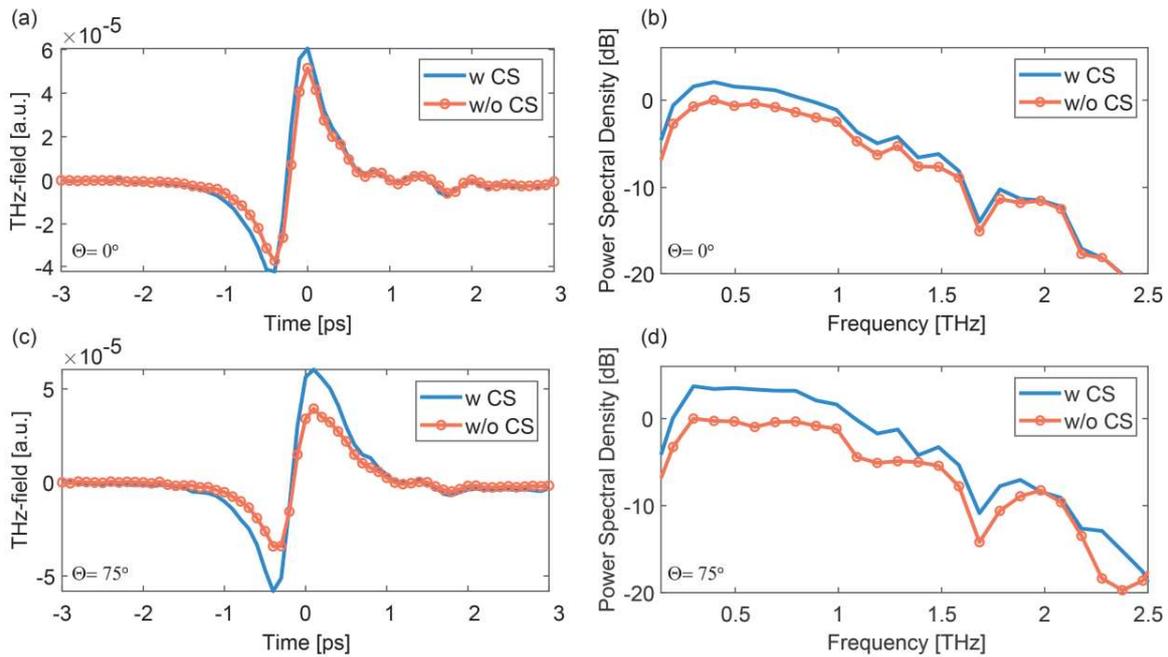

Figure 4. Experimental THz enhancement angle-dependence campaign (a, b) Time-domain trace and corresponding frequency spectrum of the THz field at normal incidence (θ = 0°); (c, d) Time-domain trace and corresponding spectrum θ=75°. All measurements were performed at a constant pump peak fluence of around 300 µJ/cm$^2$ (which also means that the overall power increases with the angle)

Representative detected THz time-domain waveforms and their corresponding frequency spectra at Θ=0° and Θ=75° are shown in Figure 4, highlighting the emission enhancement induced by the CS-layer (with an enhancement greater in the latter).

## Discussion

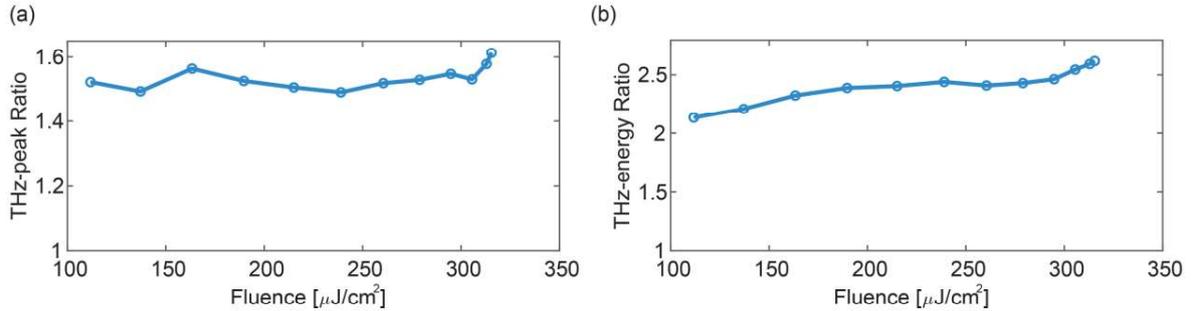

Figure 5. THz enhancement trends: (a) enhancement in THz peak field amplitude due to the CS layer, plotted as the ratio of peak fields with and without CS, as a function of pump peak fluence. (b) Enhancement in THz pulse energy, shown as the ratio of pulse energies with and without CS, versus pump peak fluence. All measurements were taken at θ=75°.

Figure 5 illustrates the dependence on the fluence of the enhancement induced by the CS layer. The ratio of THz peak field amplitudes with and without the CS layer increases with pump fluence, reaching up to 1.6 in the explored range (Fig. 5a). The corresponding trend for THz pulse energies is shown in Fig. 5b, enhancing consistently across the range, following an approximately quadratic relationship to the field enhancement. The plasmonic CS layer provides a substantial boost to the overall conversion efficiency. Notably, the fluctuations observed are primarily attributed to airflows in our experimental setup, as temperature variations in the spintronic stack substrate can perturbatively affect the conversion efficiency (see Supplementary Information).

Interestingly, at the explored fluence levels, saturation effects become significant, causing the response to deviate from a purely quadratic dependence as a result of both local saturation and thermal derating (i.e., reduced efficiency at elevated temperatures). Nevertheless, the observed enhancement persists as the excitation increases and is more pronounced at Θ=75°.

## Conclusion

Given the considerable attention devoted in the literature to the optimisation of spintronic stacks, we believe nanodecoration with CS is potentially enabling in the exploitation of STEs, with a very limited technological overhead. Although extensive quantitative benchmarks are beyond the purpose of this work, these observations suggest that the optimisations toward structures with surface nano-resonator-mediated coupling could be more rewarding than simply maximising the emission from the bare stack. Interestingly, our SEM images (see supplementary information) suggest that this coverage and nano-shell clustering (typical of dense layers) are not a significant hindering element in the enhancement. In essence, CS-enhanced spintronic emitters provide access to local efficiencies largely exceeding that of the native structure.

In conclusion, we show that the combination of plasmonic core-shells with a thin spintronic platform affects the overall conversion efficiency of STEs. By depositing a sparse monolayer of Au–$SiO_2$ CS nano-resonators on the spintronic layer, the nanoparticles act as ultrafast plasmonic couplers leading to a dramatic increase in the near-field coupling within the FM-layer, promoting terahertz generation. The macroscopic THz field enhancement achieved with very low surface coverage suggests very high local conversion enhancement. We believe this represents a key step toward new optimisation pathways in STE technology.

## Materials and methods

The sample layout consists of a core–shell nanoparticle positioned on a spintronic trilayer composed of tungsten, iron, and platinum, each with a thickness of 2 nm, deposited on a 1 mm thick $SiO_2$ substrate - BOROFLOAT® 33 (NEXTERION® Glass B) - and deposited using DC-magnetron sputtering or e-beam sputtering under a working gas pressure below $7*10^{-7}$ Torr. Details of the fabrication process are provided in the Supplementary Information.

The nanoparticles are randomly distributed across the surface. The SEM characterisation (Fig. 1b and Fig. S4) shows a general monolayer-clustered distribution of CS with an average relative surface coverage of approximately 6 % (see table in Supplementary Information for a more detailed analysis of this estimation). The CSs (from nanoComposix Inc.) have a gold-shell silica-core structure with a diameter of 154±7 nm (a silica core of diameter 117±7 nm with a gold shell of thickness 19 nm) and features a peak absorption centred at around $\lambda$=800 nm (Fig. 1b), compatible with our Ti:Sa laser pump source (data from nanoComposix Inc.). The deposition is performed via drop-casting of a dispersion in a solution of concentration 0.05 mg/mL.

## Funding declarations

This project received funding from the European Research Council (ERC) under the European Union's Horizon 2020 Research and Innovation Programme Grant No. 725046. The authors acknowledge financial support from the (UK) Engineering and Physical Sciences Research Council (EPSRC), Grant Nos. EP/Z533178 and EP/X012689/1, the Leverhulme Trust (Research Project Grant number RPG-2022-090, Early Career Fellowship ECF-2020-537, Early Career Fellowship ECF-2022-710, Early Career Fellowship ECF-2023-315, Early Career Fellowship ECF-2024-529), Loughborough University's Vice-Chancellor Independent Research Fellowship, and DEVCOM US Army Research Office, Grant agreement W911NF2310313.

## Author Contributions

V.C. managed the experimental campaign, analysed the data, and performed the nanoparticle preparation, their deposition, and took the SEM images. V.C. and A.D.T. performed the measurements reported. J.T.W. performed the simulations. A.D.T. and C.L. fabricated some of the spintronic test samples. C.L. performed the magnetic characterisation of the spintronics. V.C., A.D.T., J.T.W., CL., A.D., A.C., A.P., L.P., L.O., E.I., J.S.T.G., A.P. and M.P. contributed to the physical discussion, the assessment of the result, and the drafting of the paper. M.P. conceptualised the core research idea and physical model and led the overall research activities.

**Correspondence** and requests for materials should be addressed to Prof Marco Peccianti - *Emergent Photonics Research Centre, Department of Physics, School of Science, Loughborough University, LE11 3TU, UK*, **ORCID https://orcid.org/0000-0001-8894-496X**, Email: m.peccianti@lboro.ac.uk

**Competing interests**

The authors declare no competing interests.

## Data availability

All data required to reproduce the results published in the main text are available via Figshare (https://doi.org/10.6084/m9.figshare.29518838).

# Supplementary Material: Terahertz Emission from Spintronic Stack Nanodecorated with Drop-Cast Core–Shell Plasmonic Nanoparticles

Vittorio Cecconi,[1,2] Akash Dominic Thomas,[1] Ji Tong Wang,[1] Cheng-Han Lin,[1] Anoop Dhoot,[1] Antonio Cutrona,[1] Abhishek Paul,[1] Luke Peters,[1] Luana Olivieri,[1] Elchin Isgandarov,[1] Juan Sebastian Totero Gongora,[1] Alessia Pasquazi,[1] and Marco Peccianti[1]

[1]*Emergent Photonics Research Centre, Department of Physics, School of Science, Loughborough University, LE11 3TU, UK*
[2]*DIET, Sapienza University of Rome, Via Eudossiana 18, 00184 Rome, Italy*

## SPINTRONIC CHARACTERISATION

**Spintronic Benchmark against < 110 >-ZnTe emitter**

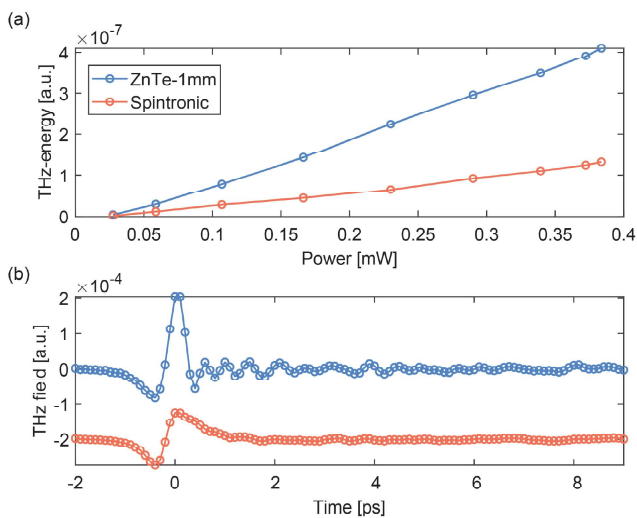

FIG. S1. Benchmark of THz emission from the native spintronic sample against a <110>-oriented 1-mm-thick ZnTe crystal. (a) Emitted THz energy as a function of pump power for both the spintronic emitter and ZnTe reference. The spintronic emission reaches approximately one-fourth of the energy of the ZnTe crystal, comparable to a ∼0.5-mm-thick ZnTe emitter. (b)Time-domain THz waveforms acquired from each emitter at pumping power $P = 166$ mW .

To quantitatively benchmark the performance of our spintronic terahertz (THz) emitter, we directly compared its emission characteristics to those of a standard <110>-oriented 1-mm-thick ZnTe crystal under identical excitation conditions. As shown in Fig. S1a, the THz energy emitted by the spintronic sample is approximately one-fourth of that measured from the ZnTe crystal across the explored range of pump powers. This suggests that the emission efficiency of our spintronic sample is comparable to that of a ∼0.5-mm-thick ZnTe crystal, with the efficiency of the core-shell enhanced sample quite exceeding this level. Figure S1b presents the corresponding time-domain THz waveforms.

**Vibrating sample magnetometry measurement**

To determine the magnetic field required to fully saturate the spintronic layer, we performed Vibrating Sample Magnetometry (VSM) on the same platform used in our core–shell experiments. The resulting VSM data provide the magnetic response of the sample, measuring the magnetisation $M$ as a function of the applied field $H$. By sweeping $H$ from positive to negative saturation, we observed a well-defined hysteresis loop (see Fig. S2) with a coercivity of approximately $7.5$, kA/m and saturation achieved for fields above approximately $20$, kA/m. Based on this result, we used Neodymium N42 permanent magnets in our experimental setup to generate an in-plane magnetic field exceeding 30mT.

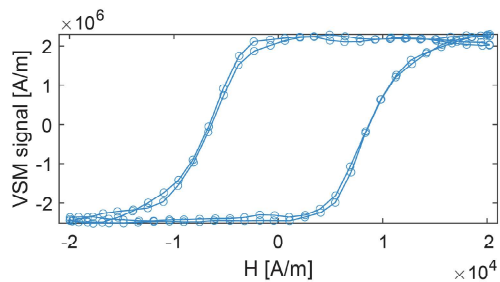

FIG. S2. VSM measurement of the spintronic sample.

**X-ray reflectivity measurement**

The layered-structure of the W/Fe/Pt spintronic emitter was characterised using a Siemens D5000 x-ray diffractometer with a Cu K$\alpha$ source (0.15418 nm). X-ray reflectivity ($\theta/2\theta$) measurements were conducted by scanning the spintronic film at incident angles ranging from 0.4° to 10°, with a step size of 0.01°. The operation voltage of 40 mV and current of 30 mA were set at two scanning ranges of 0.4° to 1° and 2° to 10°, while 30 mV and 10 mA were at 1° to 2°. Structural parameters of the W/Fe/Pt tri-layer, including thickness $t$, RMS roughness $\sigma$, and density $\rho$, were extracted by fitting the reflectivity intensity curve using the GenX software [1]. The sub-strate/W/Fe/Pt structure was first modelled with reference density values. The derived parameters for the individual layers in the tri-layer spintronic emitter are shown in Table S1. The corresponding XRR data plot is presented in Fig.S3a and SLD profile in Fig.S3b.

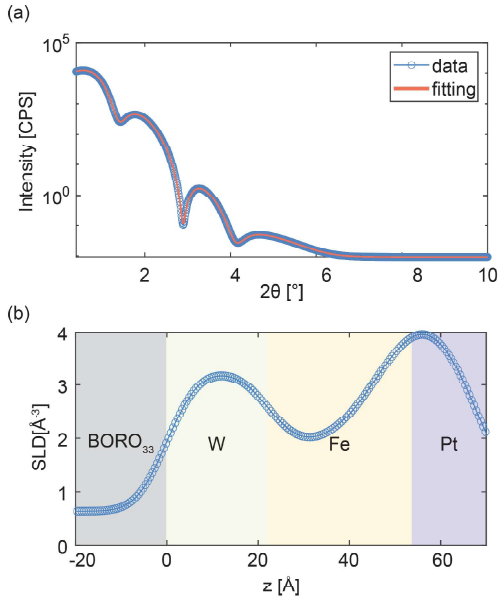

FIG. S3. (a) XRR data (red circles) with the simulated fit (black line) for the W/Fe/Pt spintronic trilayer. (b) Corresponding scattering length density (SLD) profile illustrating the depth-dependent structure of the spintronic trilayer.

TABLE S1. XRR fitting parameters for the W/Fe/Pt spintronic tri-layer were obtained using GenX reflectivity software.

| Layer | Thickness $t$ (nm) | Density $\sigma$ (atoms/Å$^{-3}$) | Roughness $\rho$ (nm) |
|---|---|---|---|
| Pt | 1.64 | 0.0828 | 1.03 |
| Fe | 2.74 | 0.0735 | 0.994 |
| W | 2.25 | 0.0473 | 0.540 |
| BOROFLOAT® 33 | $1 \times 10^6$ | 0.021 | 0.533 |

## SPINTRONIC THERMALISATION DYNAMICS

We investigate the thermal response of the spintronic emitter under optical pumping and its influence on THz emission. The results, shown in Fig.S4, present a time-resolved evolution of the THz peak amplitude as the sample transitions from a cold to a thermally equilibrated state. Measurements were conducted at both normal incidence (0°, Fig.S4b) and oblique incidence (75°, Fig.S4a). We observe that the THz emission stabilises after a few minutes, indicating that the sample has reached thermal equilibrium. Due to the variation in pump fluence between the two angles, the difference in THz peak amplitude between the cold and thermalised states is more pronounced at 0° incidence than at 75°, where the effective fluence is lower.

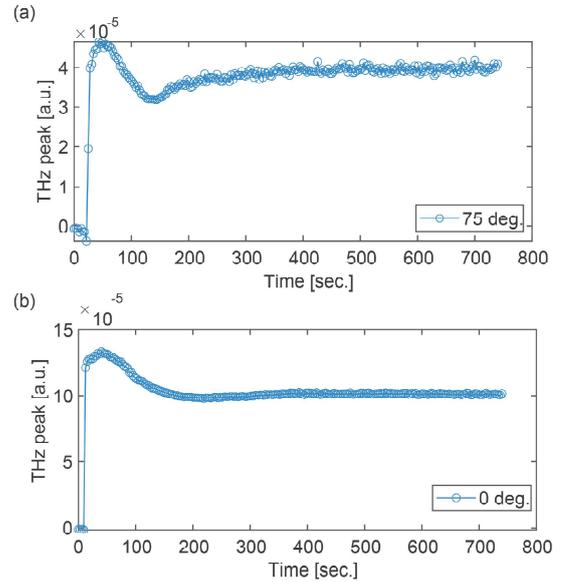

FIG. S4. Spintronic Emission vs illumination time at (a) $\Theta = 0°$ and (b) $\Theta = 75°$ at power 390mW

## NANOPARTICLES COVERAGE

We analysed multiple Scanning Electron Microscope (SEM) images to determine the distribution of nanoparticles present in each image and extracted the mean coverage using K-Means clustering. K-Means [2] is an unsupervised clustering algorithm that groups similar data points based on intensity values. It segments the image into two clusters: the background pixels (namely, the lower intensity values), and the nanoparticle pixels (with higher intensity values).

To determine the cluster centres, the algorithm calculates the mean intensity of each cluster. A threshold is then set at the midpoint between these two means to classify pixels as either nanoparticles or the background. The method then counts the white pixels (nanoparticles) and computes the percentage coverage using the formula:

$$\text{Coverage} = \frac{\text{Nanoparticle Pixels}}{\text{Total Pixels}} \times 100$$

This method allows us to efficiently and consistently quantify nanoparticle coverage across a large set of SEM images. By analysing 29 images, we determined that the average surface coverage is 6±3 % (see Fig.S5), indicating a relatively sparse distribution of nanoparticles with spatially distributed plasmonic 'hot spots'. These findings suggest that the significant THz field enhancement ob-served in our experiments arises from exceptionally high local efficiency within the plasmonic regions.

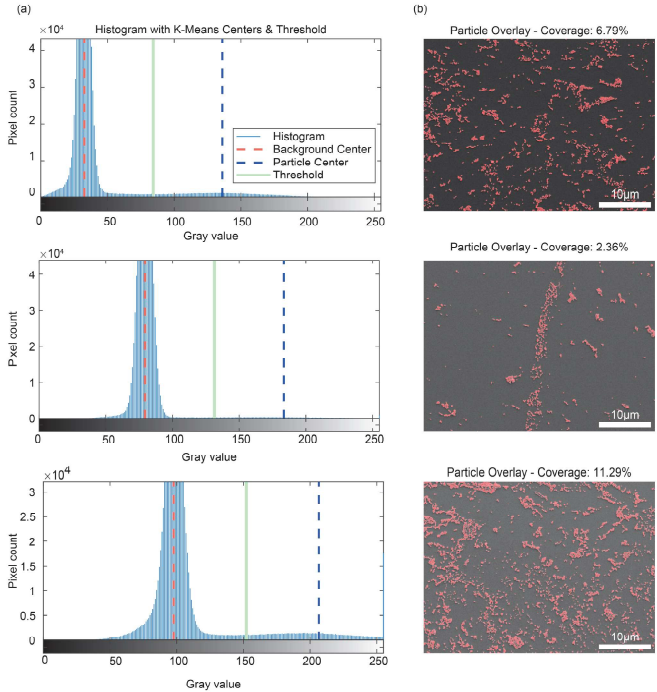

FIG. S5. (a) Histogram of grayscale intensities with K-Means cluster centres (background and particle) and the computed threshold indicated. (b) Overlay of detected nanoparticles (in red) on the original image, highlighting the area coverage percentage and the almost perfect identification of the nanoparticle coverage. SEM images of the nanoparticles deposited on the surface of the spintronic sample reveal a notably sparse distribution, with the majority forming a monolayer.

## MULTIPOLE DECOMPOSITION OF PLASMONIC RESPONSE

The numerical simulations of the core–shell nanoparticle presented in Figs. 2-3 of the main text reveal giant near-field resonant effects arising from localised surface plasmon resonances. To gain deeper insight into the optical properties of a single core–shell nanoparticle, we analyse the origin of its resonant features by performing a multipole expansion under Cartesian coordinates [3].

Specifically, we start by expressing the relation between the local electric field and the induced current density within the nanoparticle as $\mathbf{J} = i\omega\epsilon_0 \left(\tilde{\epsilon}_r - 1\right) \mathbf{E}$, where $\omega$ is the angular frequency, $\epsilon_0$ is the free-space permittivity, and $\tilde{\epsilon}_r$ is the complex relative permittivity describing both the core and gold shell. We can then enumerate the different multipole moment contributions as:

**Electric dipole moment:**

$$\mathcal{P} = \frac{1}{i\omega} \int d^3r\, \mathbf{J} \tag{S.1}$$

**Magnetic dipole moment:**

$$\mathcal{M} = \frac{1}{2c} \int d^3r\, (\mathbf{r} \times \mathbf{J}) \tag{S.2}$$

**Toroidal dipole moment:**

$$\mathcal{T} = \frac{1}{10c} \int d^3r\, \left[(\mathbf{r} \cdot \mathbf{J})\mathbf{r} - 2r^2\mathbf{J}\right] \tag{S.3}$$

**Electric quadrupole moment:**

$$Q^{(e)}_{\alpha\beta} = \frac{1}{i2\omega} \int d^3r\, \left[r_\alpha J_\beta + r_\beta J_\alpha - \frac{2}{3}(\mathbf{r}\cdot\mathbf{J})\delta_{\alpha\beta}\right] \tag{S.4}$$

**Magnetic quadrupole moment:**

$$Q^{(m)}_{\alpha\beta} = \frac{1}{3c} \int d^3r\, \left[(\mathbf{r}\times\mathbf{J})_\alpha r_\beta + (\mathbf{r}\times\mathbf{J})_\beta r_\alpha\right] \tag{S.5}$$

where $c$ is the speed of light in vacuum and $\alpha, \beta = x, y, z$.

The total scattered power into the far-field is the sum of the contribution from all multipoles as

$$P = P_\mathcal{P} + P_\mathcal{M} + P_\mathcal{T} + P_{Q^{(e)}} + P_{Q^{(m)}}$$

where,

$$P_\mathcal{P} = \frac{\omega^4}{12\pi c^3}|\mathcal{P}|^2 \quad P_\mathcal{M} = \frac{\omega^4}{12\pi c^3}|\mathcal{M}|^2 \quad P_\mathcal{T} = \frac{\omega^6}{12\pi c^5}|\mathcal{T}|^2$$

$$P_{Q^{(e)}} = \frac{\omega^6}{160\pi c^5}\sum_{\alpha\beta}|Q^{(e)}_{\alpha\beta}|^2 \quad P_{Q^{(m)}} = \frac{\omega^6}{160\pi c^5}\sum_{\alpha\beta}|Q^{(m)}_{\alpha\beta}|^2$$

The core-shell nanoparticle is numerically modelled and simulated using finite-element method as a silica core [4] with radius of 58 nm and gold shell with thickness of 19 nm, consistent with the experimental setting. The spintronic structure is modelled as a multilayer with 2 nm platinum, 2 nm iron, and 2 nm tungsten [5]. Consistently with our analysis in Fig. 2 of the main text and with the experimental procedure, we excite the particle with a linearly polarised, normally incident plane wave. As shown in Fig.S6, the multipole analysis reveals that the electric dipole response dominates the scattering properties of the core-shell nanoparticle over a wide wavelength range, spanning from 500 nm to 1000 nm. The field-superposition of all the responses conflates in the cross-section "1 CS" in Fig. 2d, for an isolated nanoshell. Fig. 2a evidences the typical features of the electric-dipole type resonances, characterised by two clear hot spots along the field polarisation axis around the nanoparticle boundary with dipole direction aligned with the incident field polarisation.

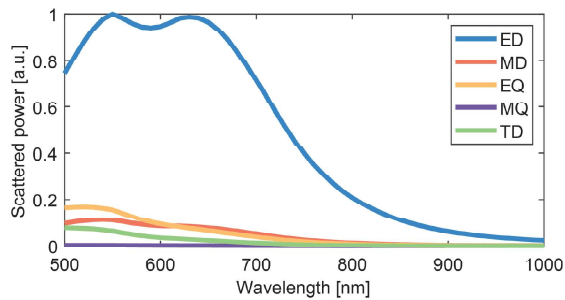

FIG. S6. Numerical characterization of optical properties of core-shell nanoparticle calculated scattered power for the dominant electromagnetic multipoles induced in the nanoparticle. ED: electric dipole; MD: magnetic dipole; TD: toroidal dipole; EQ: electric quadrupole; MQ: magnetic quadrupole.

## EXPERIMENTAL SETUP

In Figure S7 we show the experimental setup schematic for a terahertz time-domain spectroscopy [6–9]. The excitation pulses are supplied by a mJ-class Ti:Sapphire regenerative laser (Coherent Libra-HE) generating 76 fs pulses centred at $\lambda = 800$ nm with a 1 kHz repetition rate. The beam diameter (intensity at $1/e^2$) is $\approx$9mm. The setup comprises two separate beamlines, the THz excitation pump, and the optical sampling probe for detection. By placing a cylindrical neodymium magnet on the side of the spintronic sample, we generate a static magnetic field parallel to the film plane, and therefore, we generate the THz pulse by impinging the femtosecond pulse on the surface of the sample. THz electro-optic detection is implemented by co-propagating the THz field and the optical sampling probe in a standard 1 mm thick <110> ZnTe crystal [10]. The emitted THz wave is measured through a f-f (Fourier) condition on the detection crystal. The detection crystal and the probe polarisation are set to detect the p-polarised THz field. As in standard time-domain spectroscopy schemes, the time-domain traces are reconstructed by varying the delay $t_d$ between the THz pulse and the optical probe. The THz signal is measured with a balanced photodetection unit feeding a lock-in amplifier.

## SUPPLEMENTARY DATA AVAILABILITY

Data of the supplementary figures are available from the corresponding author upon reasonable request.

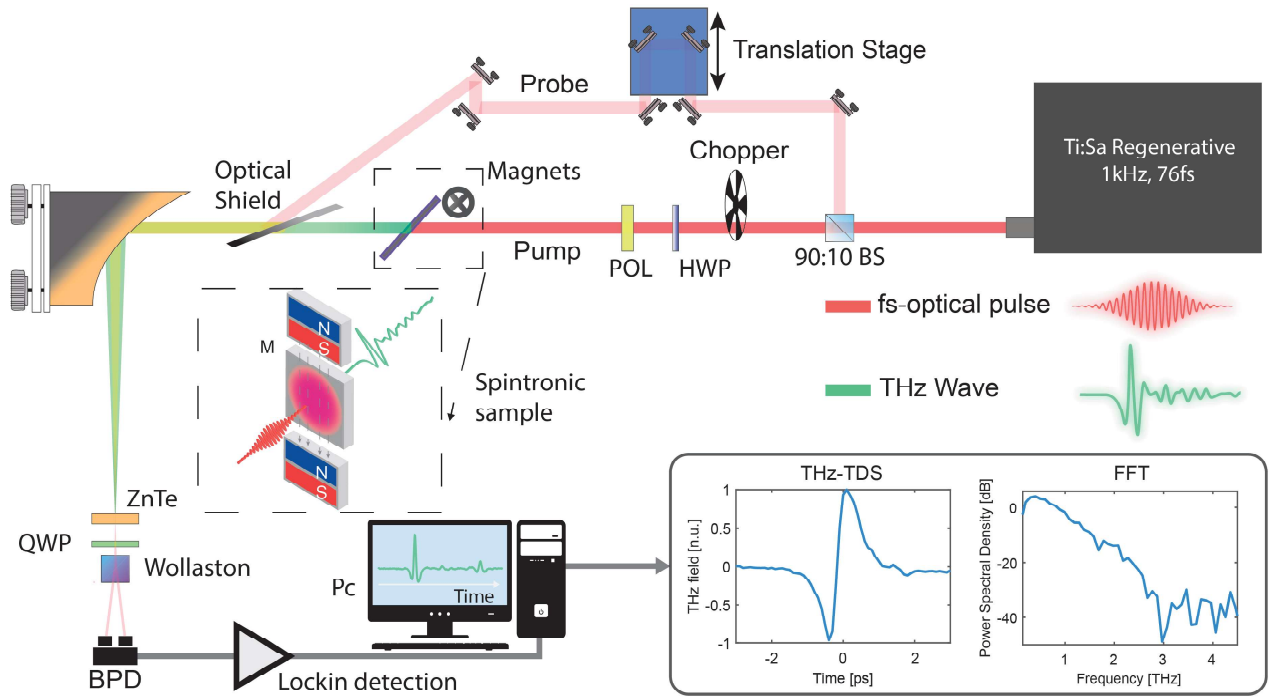

FIG. S7. Scheme of the time-domain spectroscopy apparatus for measuring the ultrafast THz pulses. BS: beam sampler, HWP: half wave-plate, QWP: quarter wave-plate, POL: polariser, ZnTe: zinc telluride, and BPD: balanced photodetectors.